\documentstyle [12pt]{article}
\newcommand{\bea}{\begin{eqnarray}}
\newcommand{\eea}{\end{eqnarray}}
\newcommand{\be}{\begin{equation}}
\newcommand{\ee}{\end{equation}}
\newcommand{\nn}{\nonumber}

\begin{document}
\def\refname{References}
V.A.Meshcheryakov  and D.V.Meshcheryakov

\begin{center}
{\Large\bf RIEMANN SURFACES OF SOME STATIC ISPERSION MODELS
AND PROJECTIVE SPACES } \\

\bigskip

\end{center}

 The S-matrix in the static limit of a dispersion relation has a finite
order N and is a matrix of meromorfic functions of energy $\omega$ in the
plane with cuts $(-\infty,-1],[+1,+\infty)$. In the elastic case it reduces to
N functions $S_{i}(\omega)$  connected by the crossing symmetry matrix A.  The
problem of analytical continuation of $S_{i}(\omega)$ from the physical sheet
to unphysical ones can be treated as a nonlinear system of difference
equations. It is shown that a global analisis of this system can be carried
out effectively in projective spaces $P_{N}$ and $P_{N+1}$. The connection
between spases $P_{N}$ and $P_{N+1}$ is discussed.

\section{Introduction}

The low-nergy hadron scattering problem remains in the focus of attention [1].
The successful development of QCD poses the question of the validity
of the analytic properties of hadron-hadron process amplitudes
previously proved for strong interactions.In a series of works by
Oehme [2], it was recently shown that they remain valid in QCD as
well.We consider the nonrelativistic limit of the dispersion relations,
which is known as static equations [3],and confine ourselves to
studying the equations of this type by reducing them to a nonlinear
boundary-value problem [4].It has the form of a series of conditions
on the S -matrix elements $S_i$.

Conditions 1.

\bea \label{} \left.  \begin{array}{l} A)~ S_i(z) - \mbox{are
meromorphic functions in the complex z plan with the
cuts}\\ ~~~~~~~~~~~~~~\mbox{with the cuts} ~
(-\infty,-1],[+1,+\infty), \mbox{i.e., the only singularities } \\
~~~~~~~~~~~~~\mbox{ of these functions in this domain are their
zeros and poles.}\\ B)~ S_i^*(z)=S_i(z^*), \\ C)~ \mid S_i(\omega+i0)\mid^2=1~
\mbox{for}~ \omega\geq 1~~
 S_i(\omega+i0)=\lim_{\epsilon\to+0}S_i(\omega+i\epsilon),\\
 D)~ S_i(-z)=\sum_{j=1}^{N}A_{ij}S_j(z).
\end{array}  \right.
\eea

  The real values of the
variable $z$ are the total energy  $\omega$ of a relativistic particle
scattered by a fixed center. The meromorphy requirement for the
functions $S_i(z)$ arises as a consequence of the static limit of the
scattering problem [5]. Elastic unitarity condition 1C holds only on
the right cut in the $z$ plane. On the left cut, the functions $S_i(z)$
are determined by crossing-symmetry conditions 1D. The
crossing-symmetry matrix A is determined by the group that leaves
the S -matrix invariant; the matrix A is known for some groups [4].
The aim in this paper is to formulate a method for studying the
Riemann surfaces of some static dispersion models.

\section{Analytic continuation of the $S$-matrix \protect \\
to nonphysical sheets}

We write Conditions 1 in a matrix form. For this, we introduce the
column $$~~S^{(0)}(z)=[S_1(z),S_2(z),\cdots,S_N(z)]^T~,$$
where the upper index denotes the physical sheet of the $S$ -matrix
Riemann surface. Conditions 1A,1B, and 1D hold on the physical
sheet, and unitarity condition 1C can be extended to the complex
values of $\omega$ , and,just like condition 1C, the extension has
the component form
$$S_i^{(0)}(z)S_i^{(1)}(z)=1$$
and analytically
continues the $S$ -matrix to the first nonphysical sheet of the
Riemann surface.To rewrite unitarity conditions 1C in the matrix
form,we introduce the nonlinear inversion transformation $I$ by the
formula $$IS(z)=[1/S_1(z),1/S_2(z),\cdots,1/S_N(z)]~.$$  As a
result, Conditions 1 take the following form.

Conditions 2.
 \bea \label{} \left.  \begin{array}{l} A)~
S^{(0)}(z) - \mbox{is a column of N meromorphic functions in the
complex} \\ \mbox{plane}~ z~ \mbox{with the cuts}~~
(-\infty,-1],[+1,+\infty), \mbox{,i.e.,the only singularities}
\\ \mbox{ of these functions in this
domain are their zeros and poles. }\\
B)~{S^{(0)}}^*(z)=S^{(0)}(z^*),\\ C)~ S^{(1)}(z)=IS^{(0)}(z),
\\ D)~ S^{(0)}(-z)=AS^{(0)}(z). \end{array} \right.  \eea

  We define the
analytic continuation to nonphysical sheets as
\be
S^{(p)}(z)=(IA)^pS^{(0)}(z(-1)^p).  \ee

 By definition (3),unitarity condition 2C and
crossing-symmetry condition 2D are easily extended to non-physical
sheets:
 \be IS^{(p)}(z)=S^{(1-p)}(z),
AS^{(p)}(z)=S^{(-p)}(-z), \ee
and we have the formula
\be
(IA)^qS^{(p)}(z)=S^{(q+p)}(z(-1)^q).  \ee
Definition (3) is motivated by the well-known solution [5] of the
problem defined by Conditions 1 for the two-row matrix

$$
A=\frac{1}{3}\left(\begin{array}{rc}-1&4\\2&1\end{array}\right).  $$
This solution for the $S$-matrix $S(z)$
is given by
\be S(z)=\left(\begin{array}{rc}W(W-2)/(W^2-1)\\W(W+1)/(W^2-1)\end{array}\right)D(z), \ee
where
$W=w+i\sqrt{z^2-1}\beta(z),~ w=1/\pi \arcsin z,~ \beta(z)=-\beta(-z)$
is a meromorphic function,and $D(z)=D(-z) $
is the Blaschke function of the variable $\zeta=\frac{1+i\sqrt{z^2-1}}{z}$.
The Blaschke function is given by
$$D(\zeta[z])=\zeta^{\lambda}\prod_n \frac{|\zeta_n|}{\zeta_n}
\frac{\zeta_n-\zeta}{1-\zeta_n^* \zeta} $$
where $\lambda$
is the order of zero, and the set of zeros
$\{\zeta_n\}$ , $|\zeta_n|<1$,
is symmetric with respect to the origin
and the axes $Im \zeta =0, Re \zeta=0$.In addition to solution
(6), Conditions 1 allow a trivial solution: the column of identical
Blaschke functions
$$
 S(z)=\left(\begin{array}{rc}1\\1\end{array}\right)D(z).  $$

Conditions 2 therefore do not determine the form of the Riemann
surface of  $S(z)$ uniquely. For solution (6), the Riemann surface of
$S(z)$ is infinite-sheeted because of the function $w$, and the
equalities
$$S^{(0)}(z)=S(W)_{|w|\leq 1/2},~ S^{(\pm n)}(z(-1)^{(\pm
n)})=S(W)_{|w\pm n|\leq 1/2},$$
hold, which allow rewriting Eqs.(5) as
\bea && (IA)^nS(W)=S(W+n),\\ &&
(AI)^nS(W)=S(W-n).  \nn \eea

Equations (7) are a system of nonlinear autonomous difference
equations and can naturally be called the dynamic form of the static
dispersion relations. The same term can therefore be used for
Eqs.(5) as well. Unlike Eqs.(7), they form a system of nonlinear
difference equations in which the number of a sheet of the Riemann
surface serves as an argument and the energy variable $z$ is a
parameter.

\section{ Formulation of the problem \protect\\ in projective spaces}

The example of two-row solution (6) shows that, in general, the
solution of the problem defined by Conditions 1 is determined by
$N +1$ entire functions, among which $N$ functions satisfy
crossing-symmetry condition 1D and the last one is symmetric with
respect to $z$ and
ensures the validity of unitarity condition 1C. Conditions 1A, 1B,
and 1D are homogeneous and can be considered in the projective spaces
$P_{N-1}$  and $P_N$. We define the nonlinear inversion
transformation $I_p$ such that it is correct in these spaces [6]:
$I_p$, где $$I_p={\Pi}_{j=1,i\not=j}^mS_j,$$ $$  m=N-1,N~.$$
We reformulate the problem defined by
Conditions 1 for these spaces. For the space $P_{N-1}$ ,the
crossing-symmetry matrix has the form specified by Conditions 1; for
the space $P_{N}$ ,its dimensionality increases by one, i.e. ,
$$A_{N-1}=A,
A_N=\left(\begin{array}{cc}A&0\\0&1\end{array}\right),$$
where $A_N$ is a block matrix. As a result, instead of
Conditions 1, we obtain the following set of requirements on a column
of $m$ functions.

Conditions 3.

\bea \label{}
\left. \begin{array}{l}
A)~ S^{(o)}(z) - \mbox{is a column of}~ m~ \mbox{meromorphic
functions in the complex}\\ \mbox{z plane~} \mbox{with the cuts}~
(-\infty,-1],[+1,+\infty), \mbox{i.e., the only singularities}
\\ \mbox{of these functions in this domain are their zeros and
poles}\\ B)~ {S^{(0)}}^*(z)=S^{(0)}(z^*),\\ C)~
S^{(1)}(z)=I_pS^{(0)}(z),\\ D) S^{(0)}(-z)=A_mS^{(0)}(z).
\end{array}\right.  \eea

We illustrate the scheme of the solution for two-row case in
terms of the projective spaces $P_1,P_2$.
We let $(x_o,x_1)=(S_1,S_2)$
denote the coordinates of the point $x$ in the space $P_1$.
We introduce the affine coordinate $X=x_{0}/x_{1}$ on the projective
line $P_1$. Setting $z =0$ in (3), we obtain the law for continuing
the coordinate $X^{(0)}$ from the physical sheet to the first
nonphysical sheet:
 \be       X^{(1)}=\frac{2X^{(0)}+1}{-X^{(0)}+4}.  \ee
Taking the nth
power of linear fractional transformation (8) and using
crossing-symmetry condition 3D, we find that
\be
X^{(0)}=-2~ и ~X^{(n)}=\frac{n-2}{n+1}.  \ee

On of crossing-symmetry conditions 3D thus proves unnecessary. This
conclusion remains valid for $3\times 3$ crossing-symmetry matrices.  The
solution of the two-row problem for the line $P_1$ allows finding
only the ratio of the functions $S_1$ and $S_2$ .The functions
themselves can be found from the solution for the projective plane
$P_2$. We write the projective coordinates of the point
$(x)=(x_0,x_1,x_2) $ in $P_2$ in a basis explicitly taking the
crossing symmetry into account:
\bea
x_0 &=& s-2a\nn\\
x_1 &=& s+a\\
x_2 &=& c,\nn
\eea
where $s$
and $c$ are symmetric functions of $z$ and $a$ is an antisymmetric
function of $z$.

Considering the transformation $(I_pA_2)^n$ in the basis $s, a, c$, we can
easily see that $s, a$, and $c$ are related by
\be
s^2-a^2-sc=0,
\ee
which is invariant under the transformations $I_p$ and $A_2$.
In other words, Eq.(12) in $P_2$ defines an invariant curve C whose
points do not leave C under the action of the transformations $I_p$
and $A_2$. In the basis $(x_0,x_1,x_2)$ ,the equation of the curve C
is given by

\be
{x^2}_1+2x_0x_1-2x_1x_2-x_0x_2=0.  \ee

Using Eqs.(10) and (13), we can easily find that
\be \frac{x_1}{x_2}=\frac{n}{n-1} \ee
and thus completely define the functions $S_1$ and $S_2$.
Taking unitarity condition 1C (which has not been used yet)into
account, we can recover formula (4) completely. We discuss the
relation between the descriptions of the two-row problem defined by
Conditions 1 for the spaces $P_1$ and $P_2$. In the projective plane
$P_2$, the solution is given by invariant curve (13). It is
irreducible and rational as is any algebraic curve of the second order.
In the affine coordinates,it becomes
$$ x=\frac{x_0}{x_2}, ~y=\frac{x_1}{x_2}, ~x^2+2xy-2x-y=0.$$

If w construct a bundle of lines of the form $\lambda_0 g_0+\lambda_1 g_1$
with the bas point $(x_0,y_0)$ in curve (13), then
the coordinates of the second intersection of the lines in the bundle
with curve (10) are rational functions of  $k=\lambda_1/\lambda_0$:
$$
x=\frac{-(x_0+2y_0)+2+k}{1+2k} , ~y=y_0+k(x-x_0).$$
The functions $x$ and $y$ are reduced to formulas (10) and (14) by
the specially chosen parametrization
$$ k=\frac{(-x_0-2y_0+1)n+x_0+2y_0-2}{n+1}~,$$
which depends on the base point of the bundle. A bundle of
lines behaves as the projective space $P_1$ under collineations
(linear transformations with nonzero determinants)in the space $P_2$.
The projective space $P_1$ is thus represented by any bundle of lines
whose base point lies on invariant curve (13) of the space $P_2$.
In [4], the invariant manifolds for the problem defined by Conditions 1
with dimensionalities $N\geq 3$ were studied and constructed using
series in a neighborhood of the rest points of dynamic systems (5).
Using projective spaces, we can reconsider this problem from a new
standpoint. We consider the problem defined by Conditions 1 with the
three-row matrix
\bea
A=\left(\begin{array}{rcc}1/3&-1&5/3\\-1/3&1/2&5/6\\1/3&1/2&1/6\end{array}
\right),
\eea

which describes the scattering of two particles whose angular momenta
are equal to unity. In the space $P_3$, the matrix $A_3$ has three
eigenvalues equal to +1 and one eigenvalue equal to -1. The
coordinates of the point $(x)$ in $P_3$ can be expressed in terms of
three symmetric functions $s_1, s_2$ and $s_3$ of $z$
and one antisymmetric function $a$ of $z$ by an ordinary collineation
(an automorphism of the projective space ):
$$x_i=b_{ij}s_j+b_{i4}a.$$
We construct a plane in $P_3$ that is invariant under the linear
transformation of the coordinates of $x$ determined by the matrix
$A_3$. It is given by
\be
 c_0x_0+c_1x_1+(2c_0+c_1)x_2+c_2x_3=0~.
\ee
It is easy to see that the plane $x_1+x_2=0$ is a particular case of
plane (16) and is invariant under the transformation $I_p$. This
plane is the space $P_2$ in which the problem defined by Conditions 1
with matrix (15) is reduced to the solvable two-row problem [7]. The
plane $x_1+x_2=0$ does not contain the rest point $\bar x=(1,1,1,1)$ of
the dynamic system defined by Conditions 3, i.e., the fixed point of
transformation (5). If we require the point $\bar x$ to lie in plane
(16), then we obtain the equation
\be
 c_0x_0+c_1x_1+(2c_0+c_1)x_2-(3c_0-2c_1)x_3=0~.
\ee

The transformation $I_p$ maps plane (17) onto
the cubic surface
\be
с_0x_1x_2x_3+c_1x_0x_2x_3+(2c_0-c_1)x_0x_1x_3-(3c_0+2x_1)x_0x_1x_2=0
\ee
in $P_3$, which is not
invariant under the transformation $A_3$.

Th intersection of plane (17) and surface (18) determines a planar
spatial curve $C$, which is not invariant under the transformation
$A_3$ in general. Indeed, excluding $x_3$ from Eqs.(17) and (18), we
obtain a third-degree homogeneous equation $G(x_0,x_1,x_2)=0$. In the
basis $s_1,s_2,a$, the function $G$ on the space $P_2$ contains odd
powers of the antisymmetric function $a$ for any $c_0$ and $c_1$. The
coefficient of $a$ is a quadratic form with respect to $s_1,s_2$,
and $a$. The invariance of the planar spatial curve $C$ under the
transformation $A_3$ implies that this quadratic form should
vanish. As any second-degree equation, it defines rational functions
$s_1,s_2$ and $a$ of some parameter $t$. Substituting them in the even
part (with respect to $a$ ) of the function $G(x_0,x_1,x_2)$, we
obtain a third-degree equation with respect to $t$, which has three
solutions in general. An invariant curve exists only if this equation
is identically zero,i.e., if $G$ is reducible. The equation
determining the coefficients $c_0$ and $c_1$ is given by
\be
R_{x_0}(G,G'_{x_1})\equiv0 \ee

where  is the resultant of $G$  and $G'_{x_1}$ with respect to  $x_0$.
From Eq.(19), we obtain $c_0=-1, c_1=3$ and
find the function
\be G(x_0,x_1,x_2)=(-3x_1^2+x_0x_1+3x_0x_2-x_1x_2)(-x_0+x_2)=0~,\ee
which defines the reducible curve $C$. The first factor in Eq.(20) is
invariant under the transformations $I_p$ and $A_2$ and together with
Eq.(17) defines the well-known solution [8] with a finite number of
poles with respect to $w$. It is represented in $P_3$ as the
intersection of the plane
\bea &&
-x_0+3x_1+x_2-3x_3=0~; \eea
and the surface
\bea && -3{x^2}_1+x_0x_1+3x_0x_2-x_1x_2=0~. \eea
Using Eq.(22) and writing Eq.(21) in the form $$
x_1x_3=x_0x_2 , $$, we can
easily verify the invariance of (21) under the transformation $I_p$.
Under the action of the transformation $A_3$, the second factor in
(20) becomes $(-x_1+x_2)$; as a result, we have the degenerate
quadratic form
$$
(-x_0+x_2)(-x_1+x_2)=0~, $$
which is invariant under
the transformations $I_p$ and $A_3$. It determines two bundles of
lines that are invariant under the transformation $I_p$ and pass into
each other under the transformation $A_3$:
$$ x_0=x_2,~ \frac{x_0}{x_1}=\frac{n+1/6}{n-7/6};~~ x_1=x_2,
~\frac{x_0}{x_1}=\frac{n-3/2}{n+1/2}
$$

\section{\bf Conclusion}

The nonlinear boundary-value problem of constructing an $N$
-dimensional (condition 1A), elastically unitary (condition 1C), and
crossing-symmetric (condition 1D) $S$ -matrix is formulated in the
projective spaces $P_{N-1}$  and $P_N$. In the space $P_{N-1}$, it
can be considered the result of projecting (ignoring unitarity
condition 1C) the initial problem defined by Conditions 1 from the
affine space $A_N$ to the projective space $P_{N-1}$. The condition
for the analytic continuation of the $S$ -matrix to nonphysical
sheets is represented as a nonlinear autonomous system of difference
equations,i.e., in the dynamic form. It can also be considered a
nonlinear transformation in the spaces $A_N$, $P_{N-1}$, and $P_{N}$ .In
particular, among its fixed points, there is a point corresponding
to the $S$ -matrix without interaction. In the neighborhood of this
point, the $S$ -matrix was studied using power series in $1/w$, which
can sometimes be summed [4]. The use of the projective space
technique allows analyzing the solutions globally, i.e., constructing
the invariant subspaces containing the solutions to be found. The
invariant subspaces are determined by functions that are homogeneous
in the projective spaces $P_{N-1}$ and $P_{N}$ but not in the affine
space $A_N$. This statement disagrees with the
conclusion in [9], according to which the invariant subspaces in the
affine space $A_N$ are also determined by homogeneous functions. The
above geometric interpretation of the boundary-value problem defined
by Conditions 1 in the projective spaces $P_{N-1}$ and $P_{N}$ and the
examples considered in [5] and [8] indicate that the homogeneity
requirement on the functions defining the invariant subspaces of
$A_N$ should be rejected.  Concrete applications of the described
procedure for solving the nonlinear boundary-value problem are
demonstrated in Appendices 1 and 2.

\newpage
\vskip 0.1 in
\centerline{\bf Appendix 1 }
\vskip 0.1 in

Th two-row crossing-symmetry matrix for the group SU(2) is given by
$$
A_2=\frac{1}{2l+1}\left(\begin{array}{rc}-1&2l+2\\2l&1\end{array}\right),~~l\in N.
$$
The matrix considered in the paper is particular case of it for
$l=1$. We give the calculation scheme for the general case. In the
projective line $P_1$, the first affine coordinate $X=x_{0}/x_{1}$ is
continued to the first nonphysical sheet according to the rule
$$
  X^{(1)}=\frac{2lX^{(0)}+1}{-X^{(0)}+(2l+2)}
$$

and together with the crossing-symmetry condition yields the value
of
$X^{(n)}~:$ \be
    X^{(n)}=\frac{n-(l+1)}{n+l},~~X^{(0)}=-(1+1/l)~. \ee
The relation $x_{0}/x_{1}$ is therefor defined on every nonphysical
sheet for $z =0$, and to construct the functions $S_1$ and $S_2$, it
suffices to find $x_{1}/x_{2}$. We let  $\varphi=x_{1}/x_{2}~.$ denote this
ratio.It is determined by the system of functional equations
\be \varphi^{(n)}\varphi^{(1-n)}=1~,\\ \ee \be
\frac{\varphi^{(n)}}{\varphi^{(-n)}}=\frac{n+l}{n-l}~,\\
\ee
which follow from the unitarity and crossing-symmetry conditions
(4) on nonphysical sheets. We use those Eqs.(4) here that were not
involved in deriving formulas (23). Equation (24) has an obvious
solution in the ring of meromorphic functions,
\be
\varphi^{(n)}=\frac{G(n)}{G(1-n)}~,\\
\ee

where $G(n)$ is an arbitrary entire function. Solution (26) can be
represented in another form,
$\log\varphi^{(n)}=g(n-1/2)~,$
where
$g(n-1/2)$ is an odd function of its argument. This form of the
function $log\varphi^{(n)}$ is convenient for solving Eq.(25), which
can be easily rewritten as
\be g(n+1)+g(n)=log\frac{n+1/2+l}{n+1/2-l}~.\\ \ee
We can find a particular solution of inhomogeneous difference
equation (27) by consecutively changing the unknown function
according to the formula
$$g_{m}(n)=g_{m+1}(n)+log\frac{n+(-1)^{m}\alpha_{m+1}}
{n-(-1)^{m}\alpha_{m+1}}~,$$
where  $\alpha_{k}=1/2+l-k$ and $g_{0}(n)=g(n)$.
The function $g_{k}(n)$ satisfies the equation
$$g_{k}(n+1)+g_{k}(n)=log\frac{n+1/2+(-1)^{k}(l-k)}{n+1/2-(-1)^{k}(l-k)}$$
and we obviously have $g_{l}(n+1)+g_{l}(n)=0$. The general solution
of this homogeneous equation determines the function $D(z)$, which
enters formula (6) and places no restrictions on the form of the
invariant constraints on $x_0,x_1,x_2$. We therefore set $g_{l}=0$
and obtain the expression for $\varphi^{(n)}$:
\be
\varphi^{(n)}=\prod\nolimits_{m=1}^{l}\frac{n-1/2-(-1)^{m}(1/2+l-m)}{n-1/2+(-1)^{m}(1/2+l-m)}~.
\ee
Excluding the parameter $n$ from Eqs.(23) and (28), we obtain an
equation determined by a homogeneous polynomial in $x_0,x_1,x_2$
of degree $l +1$; it gives Eq.(13) for $l =1$.

\vskip 0.1 in
\centerline{\bf Appendix 2}
\vskip 0.1 in

We apply the developed method to the problem of scattering a
pseudoscalar meson with unit angular momentum by a fixed nucleon with
the same angular momentum. In this case, the crossing-symmetry matrix
is given by expression (15). We decompose the column $S(z)$ into a
sum of eigenvectors of the matrix $A$:
\be \label{matrix}
S(z)=s_1(z)\left(\begin{array}{rc}
 1 \\ 1 \\ 1 \end{array}\right)+ \frac{1}{4}
 s_2(z)\left(\begin{array}{rc} 15 \\ -5 \\ 3 \end{array}\right)
 +2\psi(z)\left(\begin{array}{rc} -2 \\ -1 \\ -1 \end{array}\right)
 .  \ee

For $q=1,{~}p=0$, functional equation (5) in the limit $ z
\rightarrow \infty$ determines
the fixed (rest) points of the problem. Returning from the basis
$s_1(z),{~},s_2(z),{~}\psi(z)$ to the column $S(z)$, we have
\be \label{matrix}
S=  \pm i \left(\begin{array}{rc}
 -(2 \pm \sqrt{5}) \\ -\frac{1 \pm \sqrt{5}}{2} \\ \frac{1 \pm
\sqrt{5}}{2} \end{array}\right) .  \ee

We can see from (30) that all rest points lie in the plane
$S_2+S_3=0$.  This plane is invariant under the inversion
transformation $I$ and the crossing-symmetry transformation $A$. In
the plane $S_2+S_3=0$, three-row crossing-symmetry matrix (15) passes
into the two-row matrix $A_2$ \be \label{matrix}
A_2=\frac{1}{3}\left(\begin{array}{rc} 1 & -8 \\ -1 & -1
\end{array}\right),  \ee

and the problem is thus reduced to finding two functions $S_1(z)$ and
$S_2(z)$. Setting $z =0$ and defining $ X^{(k)}=S_1^{(k)}/S_2^{(k)} $,
where $k$ is the number of the sheet of the Riemann surface, we see
that the transition from the physical sheet to the sheet with the
number $n$ is realized by the linear fractional transformation
\be   \label{matrix}
X^{(n)}=\sqrt{5}
\frac{\sqrt{5}(X^{(0)}-2)(y_{-}^n-y_{+}^n)+
      (X^{(0)}+4)(y_{-}^n + y_{+}^n)}
       {(X^{(0)}+4)(y_{-}^n-y_{+}^n)+
       \sqrt{5}(X^{(0)}-2)(y_{-}^n + y_{+}^n)}, \ee
where $y_{\pm}=(3 \pm \sqrt{5})/2$.The
unitarity and crossing-symmetry requirements on $X^{(n)}$ give the
condition
\be (X^{(0)}-2)(X^{(0)}+4)=0 \ee

which determines $X(0)$ .Consequently, we obtain two different
solutions, $X^{(0)}=2$ and $X^{(0)}=-4$ , which are compatible with
the unitarity and crossing-symmetry requirements.

The ratio $S_{1}/S_{2}$
is thus determined for $z =0$ on every nonphysical sheet of the
Riemann surface defined by Conditions 2 with matrix (31),and to
construct $S_{1}$ and $S_2$ ,it suffices to find any of these
functions.  We set $S_2(n)=\Phi(n)=-s_2(n)+\psi(n)$, where $s_2$
and $\psi$ are the functions introduced in (29).This function
satisfies the system of functional equations
\be \Phi(1-n)\Phi(n)=1~, \ee

\be
\frac{\Phi(n)}{\Phi(-n)}=(-1) \frac{ch log y_{+}^{n+1/2}}
{ch log y_{+}^{n-1/2}}, {~}X^{(0)}=2
\ee

\be
\frac{\Phi(n)}{\Phi(-n)}=(-1) \frac{sh log y_{+}^{n+1/2}}
{sh log y_{+}^{n-1/2}}, {~}X^{(0)}=-4
\ee

Relation (32) is used in deriving Eq.(35). Equation (34) has the
solution \be \Phi(n)=e^{g(n-1/2)}, \ee, wher $g(n)$ is an arbitrary
odd function, $g(n)=-g(-n)$. Substituting (37) in (35) and changing
$n \rightarrow n+1/2$, we obtain the difference equation
\be g(n+1)+g(n)=log(-1) \frac{ch((n+1) logy_{+})}{ch(n logy_{+})},
{~~}X^{(0)}=2 \ee \be g(n+1)+g(n)=log(-1) \frac{sh((n+1)
logy_{+})}{sh(n logy_{+})}, {~~}X^{(0)}=-4 \ee
for the unknown function $g(n)$.

Solving Eq.(37) by the method of consecutive functional changes, we
obtain
\be
g(n)=g_{-1}(n)+g_{\infty}(n) +\sum_{m=0}^\infty G_m(n) \ee
where $g_{\infty}(n)=nlogy_{+}$
and
\be G_m(n)=log\frac {ch ({(n+1+2m)}log
y_{+})ch ({(n-2(m+1))}log y_{+})} {ch({(n-1-2m)} log y_{+})ch
({(n+2(m+1))} log y_{+})}, {~}X^{(0)}=2 \ee

\be G_m(n)=log\frac {sh({(n+1+2m)} log
y_{+})sh({(n-2(m+1))} log y_{+})} {sh({(n-1-2m)} log y_{+})
sh({(n+2(m+1))} log y_{+})}, {~}X^{(0)}=-4\ee

The term $g_{-1}(n)$ is introduced to take the factor -1 in
Eq.(37) into account. We set $e^{g_{-1}(n)}=\xi(n)$. The function
$\xi(n)$ solves the system of functional equations
\be
\xi(n+1)\xi(n)=-1,~~~\xi(n)\xi(-n)=1.
\ee

The general solution of this system is
expressed in terms of $\theta$-functions. We confine ourselves to
the degenerate case here,
\be
\xi(n)=tg \frac{\pi}{2}(n+\frac{1}{2})
\ee

We now use unitarity condition 1C. As a result, the function $n$
considered as a function of the complex variable $z$ solves the
boundary-value problem and is given by
\be n(z)=1/\pi\arcsin z+i\sqrt{z^2-1}\beta(z),
\ee
where $\beta(z)=-\beta(-z)$ is an arbitrary meromorphic
function. It follows from Eq.(42) that the Riemann surface of the
model under consideration has algebraic ramification points at
$z=\pm1$ and a logarithmic ramification point at infinity. Formulas
(32),(33),(37)-(39),(41), and (42) now give the general solution of
the problem defined by Conditions 1 for crossing-symmetry matrix
(31).

\newpage

\begin{center}              {\bf REFERENCES}  \end{center}
\begin{enumerate}

\item
A.L.Machovariani,A.G.Rysetsky.  Nuclear Phys.A. 1990. v.515 P.
621.

\item
R.Oehme. Phys.Rev.D 1990. V.42 P. 4209; Phys. Lett. B. 1990. V.252.
P.14; J.Mod.Phys.A. 1995. V.10. P.1995.

\item G.F.Chew.  F.E.Low. Phys.Rev. 1956. V.101 P.1570.

\item
V.A.Meshcheryakov, Statistical models in discrete approach [in
Russian ],Preprint P-2369, Joint Inst. Nucl.  Res.,Dubna
(1965); V.I.Zhuravlev and V.A.Meshcheryakov, Fiz.Elem.Chast.At.Yadra,
5, 173 (1974).

\item
G.Wanders. Nuovo
Cim. 1962. V.23 P.816;\\ V.A.Meshcheryakov, JETP, 24, 431 (1967).

\item
V.A.Meshcheryakov, Solutions of nonlinear problems of dispersion
relations in projective spaces, in:Proc.  Symp.in Ahrenshoop
(Preprint PHE81-7),Institut fur Hoohenenergiephysik, Akademie der
Wisserschaften, Zeuhen, DDR (1981),p.44.

\item
V.I.Zhuravlev, V.A.Meshcheryakov, and K.V.Rerikh, Sov.J.Nucl.Phys.,
10 ,96 (1969).

\item
V.A.Meshcheryakov, Dokl.Akad.Nauk SSSR, 174, 1054 (1967).

\item
M.Froissart,  R.Omnes. Comptes Rendus Acad.Sci. 1957. V.245 P.2203.

\end{enumerate}

\end{document}